%% file: main.tex
\documentclass[a4paper]{article}
\usepackage[margin=25mm]{geometry}

\usepackage[utf8]{inputenc} 
\usepackage[T1]{fontenc}    
\usepackage[hidelinks]{hyperref}    

\usepackage{url}            
\usepackage{amsmath}
\usepackage{booktabs}       
\usepackage{amsfonts}       
\usepackage{nicefrac}       
\usepackage{microtype}      
\usepackage{lipsum}
\usepackage{authblk}    
\usepackage{xcolor}
\usepackage{graphicx}       
\usepackage{float}
\usepackage{multirow}
\usepackage{array}
\usepackage{caption}
\usepackage[toc,page]{appendix}
\usepackage{mdframed}
\usepackage{listings}
\usepackage{xcolor}
\usepackage{nameref}
\usepackage[nolist]{acronym}  

\immediate\write18{texcount -tex -sum  \jobname.tex > \jobname.wordcount.tex}
\mdfdefinestyle{promptbox}{
    backgroundcolor=gray!10,
    roundcorner=5pt,
    linewidth=1pt,
    linecolor=gray!50,
    innerleftmargin=10pt,
    innerrightmargin=10pt,
    innertopmargin=10pt,
    innerbottommargin=10pt
}

\lstdefinelanguage{json}{
    basicstyle=\normalfont\ttfamily,
    numbers=left,
    numberstyle=\scriptsize,
    stepnumber=1,
    numbersep=8pt,
    showstringspaces=false,
    breaklines=true,
    frame=lines,
    backgroundcolor=\color{gray!5},
    literate=
     *{0}{{{\color{blue!20!black}0}}}{1}
      {1}{{{\color{blue!20!black}1}}}{1}
      {2}{{{\color{blue!20!black}2}}}{1}
      {3}{{{\color{blue!20!black}3}}}{1}
      {4}{{{\color{blue!20!black}4}}}{1}
      {5}{{{\color{blue!20!black}5}}}{1}
      {6}{{{\color{blue!20!black}6}}}{1}
      {7}{{{\color{blue!20!black}7}}}{1}
      {8}{{{\color{blue!20!black}8}}}{1}
      {9}{{{\color{blue!20!black}9}}}{1}
      {:}{{{\color{red!70!black}{:}}}}{1}
      {,}{{{\color{red!70!black}{,}}}}{1}
      {\{}{{{\color{red!70!black}{\{}}}}{1}
      {\}}{{{\color{red!70!black}{\}}}}}{1}
      {[}{{{\color{red!70!black}{[}}}}{1}
      {]}{{{\color{red!70!black}{]}}}}{1},
}

\providecommand{\keywords}[1]
{
  \small	
  \textbf{\textit{Keywords---}} #1
}

\input{Acronyms}

\title{An Evidence-Driven Analysis of Threat Information Sharing Challenges for Industrial Control Systems and Future Directions}

\author[1]{\small Adam Hahn} 
\author[2,3]{\small Rubin Krief} 
\author[1]{\small Daniel Rebori-Carretero} 
\author[2,3]{\small  Rami Puzis} 
\author[3,4]{\small Aviad Elyashar} 
\author[5]{\small Nik Urlaub}
\affil[1]{The MITRE Corporation, McLean, VA, USA}
\affil[2]{Department of Software and Information Systems Engineering, Ben-Gurion University of the Negev, Beer-Sheva, Israel}
\affil[3]{Cyber@BGU, Ben-Gurion University of the Negev, Beer-Sheva, Israel}
\affil[4]{Department of Computer Science, Shamoon College of Engineering, Beer-Sheva, Israel}
\affil[5]{National Renewable Energy Laboratory (NREL)}

\begin{document}

\maketitle
\begin{abstract}
The increasing cyber threats to critical infrastructure highlight the importance of private companies and government agencies in detecting and sharing information about threat activities. 
Although the need for improved threat information sharing is widely recognized, various technical and organizational challenges persist, hindering effective collaboration. 
In this study, we review the challenges that disturb the sharing of usable threat information to critical infrastructure operators within the \ac{ICS} domain. 
We analyze three major incidents: Stuxnet, Industroyer, and Triton. 
In addition, we perform a systematic analysis of 196 procedure examples across 79 MITRE ATT\&CK® techniques from 22 ICS-related malware families, utilizing automated natural language processing techniques to systematically extract and categorize threat observables. 
Additionally, we investigated nine recent \ac{ICS} vulnerability advisories from the \ac{CISA} Known Exploitable Vulnerability catalog. 
Our analysis identified four important limitations in the \ac{ICS} threat information sharing ecosystem: (i) the lack of coherent representation of artifacts related to \ac{ICS} adversarial techniques in information sharing language standards (e.g., \ac{STIX}); (ii) the dependence on undocumented proprietary technologies; (iii) limited technical details provided in vulnerability and threat incident reports; and (iv) the accessibility of technical details for observed adversarial techniques. 
This study aims to guide the development of future information-sharing standards, including the enhancement of the cyber-observable objects schema in \ac{STIX}, to ensure accurate representation of artifacts specific to \ac{ICS} environments.
\end{abstract}\hspace{10pt}


\acresetall 

\section{Introduction}
\label{Introcudtion}
As critical infrastructure increasingly becomes the target of cyber threats, organizations and governments must be able to detect and share information about threat actor activity. 
Numerous government reports, such as Executive Order 13691~\cite{exec_order_13691} and the Cyberspace Solarium Commission report~\cite{cyberspace_solarium_2023}, have highlighted the need for enhanced information sharing to safeguard critical infrastructure. 
Additionally, numerous government programs have been established to facilitate information sharing between the public and private sectors. 

Effective information sharing enhances collective situational awareness, enabling quicker responses to emerging threats, and reducing the risk of widespread disruption.
However, realizing these benefits requires not only the collection and dissemination of threat information but also ensuring that the shared intelligence contains sufficient technical detail and can be represented in standardized formats that enable operational use by asset owners.

While the need for improved threat information sharing is well-defined, several technical and organizational challenges persist that hinder effective sharing. 
Prior work has assessed \acp{IOC} in \ac{ICS}, but has not reviewed ATT\&CK techniques to determine what observables are required to capture attack behavior.

In this paper, we examine the challenges that hinder the sharing of actionable threat information with asset owners within the \ac{ICS} domain. 
This study leverages the MITRE ATT\&CK for ICS~\cite{mitre_attack_ics} knowledge base, which aggregates cyber threat intelligence reports across known \ac{ICS} incidents. 

Our paper highlights the key challenges by reviewing three well-known incidents: Stuxnet~\cite{stuxnet_facts_report}, Industroyer~\cite{ukraine_power_grid_analysis}, and Triton~\cite{saudi_arabia_cyberattack}. 
This analysis aims to identify the adversarial techniques employed by threat actors and to determine the information that asset owners need to implement effective detection mechanisms against these techniques.
We also explore nine recent \ac{ICS} vulnerability advisories with evidence of exploitation as documented in the \ac{CISA} Known \ac{KEV} Catalog ~\cite{cisa_kev_catalog}. 

To provide a comprehensive assessment beyond these three high-profile incidents, this study extends the analysis to a systematic examination of 196 procedure examples across 79 techniques from 22 malware families documented in the MITRE ATT\&CK for ICS framework. 
Using a \ac{LLM} to extract and categorize threat observables, we analyzed 361 total observables to understand the breadth of representation and actionability challenges across the entire knowledge base. 
This comprehensive analysis revealed that only 101 observables had full \ac{STIX} observable support, 191 had partial support, and 69 lacked any \ac{STIX} representation. 
Furthermore, only 87 observables contained actionable technical details sufficient for detection development, while 274 lacked the necessary specificity for operational use.

The primary research questions driving this study are: 
 \begin{enumerate}
     \item  What are the technical artifacts necessary to detect known \ac{ICS} adversarial techniques? Can current information-sharing standards adequately represent them? (see Section~\ref{subsec:Deep Dive into Threat Information of Three Major Attacks}). 

     \item To what extent do existing threat reports provide actionable technical details for detection development? (see Section~\ref{subsec:Observation and Analysis}).  

     \item To what extent do existing vulnerability advisories provide actionable technical details for detection development? (see Section~\ref{sec:Analysis of Vulnerability Information Sharing}). 

     \item What are the main challenges and systemic barriers preventing effective threat information sharing in \ac{ICS} environments? (see Section~\ref{sec:Challenges in Threat Information Sharing}).
 \end{enumerate}

The outcome of this analysis identifies four important limitations in the ICS threat information sharing ecosystem that impact the ability of asset owners to effectively use such information. 
Specifically, this effective information sharing is limited by
(i) the lack of coherent representation of artifacts associated with known ICS adversarial techniques in information sharing language standards (e.g., \ac{STIX}), 
(ii) the reliance on undocumented and proprietary technologies, especially network protocols without open-source parsers, 
(iii) the insufficient technical details provided in vulnerability and threat incident reports, and 
(iv) the availability of technical details across all observed adversarial techniques.

This paper intends to inform the development of future information-sharing standards in ICS, including the expansion of the \ac{SCO} schema.
Our study reviews systematically ATT\&CK techniques to determine the necessary indicators or observables that represent adversarial behavior across the full spectrum of documented \ac{ICS} incidents. 
This approach provides both depth through detailed case study analysis and breadth through a comprehensive systematic assessment of the entire ATT\&CK for ICS knowledge base.

\section{Related Work}
\label{Related Work}

Recent advancements in \ac{CTI} have underscored the complexities of securing \ac{ICS} and broader \acp{CPS}, particularly in the context of critical infrastructure. 
\acp{ICS} and \acp{CPS} often rely on proprietary and specialized protocols that are difficult to monitor and integrate with conventional cybersecurity tools. 
Krasznay et al.~\cite{krasznay2021possibilities} highlighted the critical vulnerabilities in these systems, which stem from their original design priorities focused on safety rather than cybersecurity. 
The increasing connectivity of \ac{ICS} and \ac{OT} systems has exposed these vulnerabilities to sophisticated adversaries, including state-sponsored actors, necessitating a paradigm shift in cybersecurity practices. 
The SeConSys initiative in Hungary represents a significant milestone in addressing these challenges. 
This initiative facilitates sector-specific threat intelligence sharing and detection through the development of tailored \ac{CTI} feeds, honeypots that simulate energy-sector environments, and the implementation of the \ac{NIS} Directive. 
These efforts have been instrumental in promoting localized, collaborative approaches to cybersecurity and advancing frameworks for real-time threat information sharing among key stakeholders in the energy sector.

López-Morales et al.~\cite{lopez2024securing} extended these contributions by developing innovative methodologies that advance the state-of-the-art in \ac{CTI} for \acp{CPS}. 
Their work introduces the $ICS^2$ Matrix, an enhanced version of the MITRE ATT\&CK framework tailored specifically to \acp{ICS}. 
The $ICS^2$ Matrix incorporates novel attack techniques and mitigation strategies derived from an extensive analysis of \ac{PLC}-related security research spanning 17 years. 
Additionally, the development of a high-interaction satellite honeypot provides a groundbreaking tool for simulating realistic adversary interactions, collecting data on emerging threats, and advancing the understanding of satellite-specific attack vectors. 
This work further proposes the creation of a sandbox environment for connected autonomous vehicles, enabling the simulation and analysis of cyberattacks on autonomous systems to improve resilience. 
Such contributions underscore the necessity of domain-specific \ac{CTI} methodologies that address the unique operational and security challenges inherent in different \acp{CPS}.

\subsection{Empirical Measurements and Quality Assessment of CTI Sharing}
\label{Empirical Measurements and Quality Assessment of CTI Sharing}

At the ecosystem level, empirical studies have begun to quantify the effectiveness and challenges of \ac{CTI} sharing at scale. 
Jin et al.~\cite{jin2024sharing} provided one of the first comprehensive empirical measurements of \ac{CTI} sharing effectiveness, analyzing \ac{STIX} volume, timeliness, coverage, and quality across public sources. 
Their findings reveal significant variations in feed quality and highlight the challenges that organizations encounter while operating in a shared intelligence environment. 
This empirical foundation complements broader systematic reviews that examine how organizations actually use \ac{CTI} for security decision-making~\cite{ainslie2023cyber} and identify persistent barriers to effective \ac{CTI} sharing, including governance issues, misaligned incentives, and insufficient standardization~\cite{alaeifar2024current}.

Quality and trust of shared feeds remain persistent issues across the \ac{CTI} ecosystem. 
Griffioen et al.~\cite{griffioen2019quality} conducted systematic quality evaluations of cyber threat intelligence feeds, revealing inconsistent timeliness and specificity that can undermine the operational value of shared intelligence. 
Similarly, Schaberreiter et al.~\cite{schaberreiter2019quantitative} developed quantitative frameworks for evaluating trust in \ac{CTI} quality, demonstrating that trust and quality metrics vary significantly across different sources and feed types. 
Together, these studies highlight the need for more rigorous quality assurance mechanisms in \ac{CTI} sharing initiatives.

\subsection{Standardization and Technical Implementation}
\label{Standardization and Technical Implementation}

Standardization efforts play a crucial role in making threat observables portable and actionable across different organizational contexts. 
STIX 2.1, approved as an OASIS Standard (on 10 June 2021), serves as the exchange format used across many \ac{CTI} producers and consumers~\cite{stix_v21_official}. 
However, challenges remain in effectively representing complex ICS-specific artifacts within existing \ac{STIX} objects. 
Zych and Mavroeidis~\cite{zych2022enhancing} have proposed enhancements to MITRE ATT\&CK representations by STIX, to improve filtering and prioritization tasks, while RFC 9424~\cite{rfc9424} provides important clarification on the capabilities and limitations of \acp{IOC}, reinforcing the value of higher-fidelity observables for creating durable detection capabilities.

Recent advances in natural language processing have introduced new approaches for extracting threat intelligence from unstructured reports. 
While some pipelines pursue full \ac{STIX} coverage through classical NLP techniques combined with external knowledge bases~\cite{marchiori2023stixnet}, alternative approaches rely on large language models to extract observables directly from text without requiring extensive knowledge graphs. 
These automated extraction techniques show promise for scaling \ac{CTI} processing, though systematic evaluation reveals significant limitations and failure modes. 
Alam et al.~\cite{alam2024ctibench} developed CTIBench, a comprehensive benchmark specifically designed to evaluate LLM performance in cyber threat intelligence tasks across multiple cognitive dimensions, including memorization, understanding, problem-solving, and reasoning. 
Their evaluation of five state-of-the-art models revealed concerning patterns: all models struggled significantly with tasks requiring domain expertise, with even the best-performing model (GPT-4) achieving only 71\% accuracy on foundational \ac{CTI} knowledge questions. 
More critically, their analysis identified systematic failure modes, including overestimation of threat severity scores, difficulty with out-of-knowledge cutoff information, and particularly poor performance on questions related to mitigation strategies and adversarial tools. 
These findings highlight substantial gaps in LLM reliability for cybersecurity applications.

Fieblinger et al.~\cite{fieblinger2024actionable} developed a comprehensive methodology for automating the extraction of actionable \ac{CTI} using \acp{LLM} for generating Knowledge Graphs. 
Their approach systematically evaluates multiple open-source \acp{LLM} (including Llama 2 series, Mistral 7B Instruct, and Zephyr) across different extraction techniques, such as few-shot prompt engineering, guidance frameworks, and fine-tuning to optimize the extraction of meaningful triples from \ac{CTI} texts. 
Notably, their methodology demonstrates that guidance frameworks significantly improve extraction performance beyond what is achievable with prompt engineering alone, while fine-tuning proves essential for extracting triples that adhere to specified ontologies. 
The extracted triples are subsequently used to construct knowledge graphs that provide structured, queryable representations of threat intelligence, facilitating downstream applications such as link prediction and threat analysis.

The development of domain-specific evaluation datasets has become increasingly important for assessing automated threat intelligence processing. 
Bhusal et al.~\cite{bhusal2024secure} developed the SECURE benchmark using MITRE ATT\&CK as a foundation for generating cybersecurity \ac{LLM} performance evaluation across extraction, understanding, and reasoning tasks. 
Their approach demonstrates the potential of leveraging structured threat intelligence frameworks like ATT\&CK for creating systematic evaluation datasets. 
However, the reliability of \ac{LLM}-based approaches in cybersecurity contexts remains a critical concern. 
Their comprehensive evaluation of seven state-of-the-art models revealed significant limitations, including hallucinations when models lack current information, difficulty with out-of-distribution detection, and substantial performance variations across different cybersecurity task types. 
Importantly, their findings demonstrate that while \acp{LLM} show promise for cybersecurity advisory roles, closed-source models consistently outperform open-source alternatives, particularly in problem-solving and out-of-distribution scenarios. 
These results underscore the importance of meticulous validation and human oversight when deploying \acp{LLM} for threat intelligence processing, particularly in critical infrastructure environments where inaccurate information can have severe consequences.

\paragraph{ATT\&CK Framework Application in OT Environments.}
For technique-level grounding in \ac{OT} environments, the MITRE ATT\&CK framework and its \ac{ICS} extension provide structured approaches to understanding adversary behaviors~\cite{mitre_attack_design,mitre_attack_ics_design}. 
Afenu et al.~\cite{afenu2024industrial} have leveraged ATT\&CK to validate \ac{ICS} defense mechanisms and testbed implementations. 
Choi et al.~\cite{choi2021probabilistic} generated realistic \ac{OT} attack sequences based on analysis of real-world datasets. 
These applications demonstrate the framework's utility in bridging the gap between theoretical threat models and practical security implementations.

\paragraph{D3FEND Framework Extension for OT Environments.}
Recently, MITRE extended its $\text{D3FEND}^{TM}$\cite{mitred3fend2025} cybersecurity ontology to \ac{OT}, creating a structured knowledge base for defending cyber-physical systems, including new artifacts, such as controllers, sensors, and physical process actuators. 
Yet, their additional new artifacts and techniques do not fundamentally resolve the ontological issues in D3FEND's core modeling approach, which were raised back in 2023 by Oliveira et al.~\cite{oliveira2023boosting}, such as missing concepts, semantic overload, and a systematic lack of constraints that make the model under-specified. 
D3FEND can be integrated into Cyber Threat Intelligence (CTI) exchange processes, but it does not fit the exchange formats natively in the same way MITRE ATT\&CK does, as it is an ontology for defensive techniques, not a data exchange format in and of itself. While ATT\&CK is widely supported by CTI exchange standards like STIX, D3FEND lacks native support for these formats for primary CTI exchange.

\begin{flushleft}
The existing literature highlights several recurring challenges in \ac{CTI}, including limited interoperability of detection tools, insufficient standardization of threat-sharing protocols, and a lack of actionable intelligence for specific adversarial behaviors. 
While Krasznay et al.~\cite{krasznay2021possibilities} emphasized the importance of legislative and collaborative frameworks, López-Morales et al.~\cite{lopez2024securing} focused on advancing technical methodologies to enhance the granularity and applicability of \ac{CTI}. 
Both approaches converge on the need for tailored threat intelligence that integrates sector-specific requirements into practical and effective security solutions.
Building on these foundational works, this study aims to address critical gaps in the \ac{ICS} threat information-sharing ecosystem. 
Our research specifically focuses on expanding the cyber observable objects schema within the \ac{STIX} standard to ensure the accurate representation of ICS-specific artifacts and adversarial techniques. 
Unlike prior studies that primarily concentrated on domain-specific or geographically localized advancements, this work seeks to generalize \ac{CTI} enhancements across diverse \ac{ICS} environments. 
By integrating detailed technical insights into standardized threat-sharing protocols, this study aspires to bridge the gap between theoretical threat models and actionable intelligence, thereby facilitating more effective collaboration and situational awareness across \ac{ICS} stakeholders.
\end{flushleft}

\section{Key Roles and Functions in Sharing Threat/Vulnerability Information}
\label{sec:Key Roles and Functions in Sharing Threat/Vulnerability Information}

\begin{figure}[h]
        \centering 
        \includegraphics[width=0.6\textwidth]{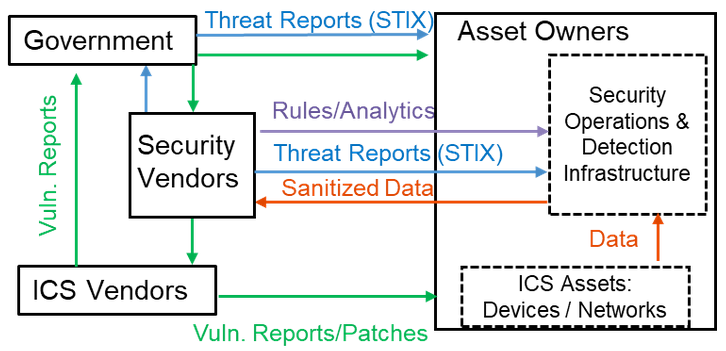}
        \caption{Overview of Threat/Vulnerability Information Sharing Roles and Technical Capabilities. Solid boxes represent stakeholders. Dashed boxes represent components. Solid arrows represent common existing information flows.}
        \label{fig:threat_vulnerability_overview}
    \end{figure}

This section provides a brief overview of the \ac{ICS} information sharing ecosystem, including key organizations, information flows, artifacts, and systems that support the sharing of threat and vulnerability information. 
Specifically, we focus on the needs of the asset owners who rely on this information to support security operations and detection efforts (see Figure~\ref{fig:threat_vulnerability_overview}). 

Typically, governments facilitate information sharing by providing reports with threat or vulnerability information to critical infrastructures and other organizations~\cite{gao_cybersecurity_2023}. 
The National \ac{CERT} is an important information-sharing hub, as it has unique access to threat information through the facilitation of information-sharing programs with both federal, defense, and intelligence agencies, as well as partnerships with private sector organizations.
   
Security vendors provide tools/platforms and information to support the detection and mitigation of cyber threats~\cite{splunk_tdir,crowdstrike_tdir}. 
Many of these vendors search for vulnerabilities and threat actor activity to develop analytics or rules to be incorporated into detection tools. 
Some vendors provide machine-readable threat information feeds, which can be directly consumed by security products, including those used to monitor for cyber threats. 
Further, many vendors obtain data from their detection infrastructures deployed at asset owner sites or collect information while supporting incident response efforts.
   
\ac{ICS} vendors typically support information sharing efforts through the disclosure of security advisories and patches when a vulnerability is discovered in their software or devices~\cite{ics_patch_tuesday}. 
Most importantly, \ac{ICS} vendors have unique knowledge regarding the technical details for their products, which are typically based on proprietary technology. 
  
Asset owners are responsible for managing the risk presented by the cyber threats. 
The larger organizations that maintain critical infrastructures typically have dedicated security operations teams, the smaller commercial asset owners commonly rely on their national \ac{CERT}~\cite{Cybersecurity_Guide_for_SMEs}. 
They may have a person partially employed in security operations to comply with regulations. 
The security team of an asset owner is responsible for monitoring their systems for threats and reporting to their auditors or the \ac{CERT}. 

Security operations teams depend on several different security products to support this, including detection platforms that monitor \ac{ICS} devices and networks for threat activities based on various detection rules and analytics. 
The security operations teams depend on governments, \ac{ICS} vendors, and security vendors to provide information about a range of vulnerabilities and threats.
  
In addition to the previously identified organizations, other entities contribute to information-sharing capabilities. 
For example, various organizations have provided open-source tools to parse network traffic, extract security-related data, and apply analytics to detect adversary behaviors shared with the organizations. 
Examples of key open-source efforts supporting the dissemination and detection of threat behaviors are included below.

\begin{itemize}
    \item \textit{Standardized threat information languages:} One key contribution to make threat information sharing more effective is the development of standard languages. 
    For example, \ac{STIX} standard provides a JSON format to communicate threat behaviors. This includes definitions of \acp{SCO}~\cite{stix_v21_official} for specific types of artifacts that can be used to detect malicious activities or collect information about them. 
    Other formats include OpenIOC\footnote{https://www.scribd.com/document/653654203/An-Introduction-to-OpenIOC} and MISP XML\footnote{https://www.misp-project.org/}.
    
    \item \textit{Detection Rules/Analytics:} Various security vendors provide rules or analytics that can be integrated within a detection platform to enable the automated detection of specific behaviors. 
    While many commercial detection platforms consume their own proprietary feed of detection rules or analytics, others rely on open-source formats. 
    Examples include the NSA ELITEWOLF\footnote{https://github.com/nsacyber/ELITEWOLF} and \ac{CISA} ATT\&CK-based Control-system Indicator Detection (ACID),\footnote{https://github.com/cisagov/ACID} and various repositories of YARA rules.
    
    \item \textit{ICS Protocol Parsers:} Many detection platforms lack inherent support for ICS protocols; therefore, dedicated parsers for \ac{ICS} protocols have been developed to enable the monitoring of ICS networks. 
    Programs like \ac{CISA}’s Industrial Control Systems Network Protocol Parsers\footnote{https://github.com/cisagov/ICSNPP} have defined several parsers for key \ac{ICS} protocols for Zeek, while Cisco Talos has also now identified Snort inspectors for specific \ac{ICS} protocols.\footnote{https://blog.talosintelligence.com/ics-protocol-coverage-snort-3/}
\end{itemize}

\textbf{Requirements for Threat Detection:}  To ensure an asset owner can detect a specific threat behavior, they must either (i) obtain a detection rule or analytic to support that detection, or (ii) develop a detection based on known technical data about that threat and the associated technology it targets. 
Accordingly, the following two requirements are defined to support the analysis of information sharing effectiveness and identify associated gaps. 
    \begin{itemize}
    \item \textit{Requirement \#1} - The information should include a detection rule/analytics or structured cyber observables that could be directly integrated within a detection platform.
    \item \textit{Requirement \#2} - The information should include sufficient technical details to enable the development of a detection rule/analytic. However, to ensure the end-user has sufficient information to establish a detection, the following additional sub-requirements must be met:
        \begin{enumerate}
        \item[a)] Technical details about the threat activity, including the adversarial technique used, and the specific artifacts manipulated during the threat activity.
        \item[b)] A thorough understanding of the underlying technology and protocols targeted by the threat is required to develop an effective detection. 
        This is particularly relevant when a threat targets a vendor's proprietary technology, especially if the technology is not adequately documented. Developing detectors for undocumented third-party technologies is a laborious and unreliable task.
        \end{enumerate}
    \end{itemize}

\section{Threat Information Sharing Analysis}
\label{sec:Threat Information Sharing Analysis}

This section reviews the procedure examples defined in MITRE ATT\&CK for ICS associated with the Stuxnet, Industroyer, and Triton events. 
While the ATT\&CK for ICS does not represent any single threat report, the procedure examples are based on information extracted from reports and provide an aggregate view of the available information about that event from numerous sources. 
Each procedure example from these three events was reviewed to extract/synthesize and document the data or technical artifacts that can be used to detect the activity. The key criteria reviewed for each procedure example are defined below.

\begin{itemize}
\item \textit{Data Source (1):} Each technique within the ATT\&CK knowledge base is mapped to a set of Data Sources that could potentially be used to detect that activity.\footnote{https://attack.mitre.org/datasources/} 
This specifies what ATT\&CK for ICS data source would have been most effective in detecting this procedure example.

\item \textit{Supporting Artifacts (2):} This more precisely defines the technical artifact that would support the detection. This identifies the specific information that needs to be shared to enable the receiving party to detect the same threat/procedure example on their network. Supporting artifacts are not explicitly defined in ATT\&CK, but they can often be derived from the procedure descriptions.

\item \textit{STIX Observable Supported (3):} This evaluates whether the artifacts needed to detect that threat are documented as \ac{SCO}. This is important because unless an artifact is documented by \ac{STIX}, it cannot be formally communicated through \ac{STIX}-based threat intelligence reports. These will be categorized into the following three types. 
While an artifact could have multiple possible \ac{STIX} observable mappings, this example only illustrates one possible mapping that supports its associated categorization.

    \begin{itemize}
    \item \textit{No Support:} \ac{STIX} lacks a relevant observable object that could correctly represent the identified artifact, thereby limiting the sharing of an observable.
    
    \item \textit{Partial Support:} \ac{STIX} has an observable that could represent this artifact, but the observable does not fully encompass the semantics of the artifact, or the observable does not define significant fields to completely represent it. 
    The partial nature implies ambiguity in the representation of observable artifacts, which in turn may yield unreliable detections or only support a limited set of aspects of the behavior. 
    For example, a network traffic object may be used to represent an ICS command, with the protocol field representing the relevant ICS protocol (e.g., \ac{SCADA}) but missing specific support to represent the command itself.
    
    \item \textit{Full Support:} An existing \ac{STIX} observable is fully capable of documenting this artifact, thereby enabling the creation of robust detection for the behavior.
    \end{itemize}
\item \textit{Artifact Detailedness level (4):} This specifies whether the information necessary to detect this specific threat behavior is available from the threat intelligence. 
This criterion is not intended to provide criticisms of the specific reports. 
Rather, it is intended to help identify whether information necessary to detect the threat is consistently available, or whether broader challenges exist in collecting the technical artifact that inhibit its sharing. 
For example, prior victims may be lacking adequate detection infrastructure or forensic capabilities to collect the evidence. 
These will be categorized into the following four types:
    \begin{itemize}
    \item\textit{Missing:} The procedure description does not mention the identified artifact nor any details about it.
    \item \textit{Mentioned:} The artifact was mentioned in the procedure description, but not addressed.
    
    \item \textit{Described:} Notable specific or distinguishing details about the artifact are described in the procedure description, but without sufficient unique information that would allow detecting it. 
    These kinds of artifacts are non-searchable.
    
    \item \textit{Actionable:} The artifact is fully described in the procedure description with unique and specific information, in a way that a detection rule can be formed to match it with low false positives. 
    These artifacts are searchable and can drive an automated response; they can be operated immediately or after a simple transform.
    \end{itemize}
    
\item \textit{Proprietary artifacts (5):} This identifies whether the artifact requires the understanding of some proprietary technology developed by a vendor/OEM, or whether that artifact is based on open/public standards or technologies. For example, many network protocols used by ICS devices lack public documentation of their functionality, while devices may contain proprietary logs or files with complex structures and content. Developing detections based on proprietary artifacts presents significant difficulty, as organizations often lack sufficient information on how to properly collect, parse, or analyze that data. These artifacts are categorized as follows:
    \begin{itemize}
    \item \textit{Open/Standard Technology:} These are technologies/protocols with publicly accessible documentation or are defined within an industry standard.
    \item \textit{Proprietary - Documented Technology:} These are also technologies/protocols developed and owned by a specific organization or company; however, in these cases, the documentation is publicly available. It is worth noting that such protocols often include undocumented features that may be relevant to the behavior of threat actors.
    \item \textit{Proprietary - Undocumented Technology:} These are technologies/protocols developed and owned by a specific organization or company. This refers to protocols for which documents are not available. The organization or company likely has internal documents, but they are not shared publicly. Proprietary but undocumented technologies can limit an organization’s ability to collect and analyze the data necessary to perform various security functions.
    \end{itemize}
\item \textit{Protocol Parsers (6):} If the artifact and associated data source include the detection of a network protocol, this identifies whether there are publicly available parsers for that protocol.
\end{itemize}

\subsection{Deep Dive into Threat Information of Three Major Attacks}
\label{subsec:Deep Dive into Threat Information of Three Major Attacks}
In this section, we review the procedure examples of Triton, Stuxnet, and Industroyer from the MITRE ATT\&CK for ICS according to the criteria defined in the previous section.

\subsubsection{Triton} 
The Triton malware~\cite{CISA2017HatMan}, also known as TRISIS or HatMan, targeted safety instrumented systems within an oil refinery by manipulating both an engineering workstation and the Triconex safety controllers used in the sulfur recovery units and burn management systems. 
While the broader campaign during this attack included activities used to gain access to the facility (references as TEMP.Veles in ATT\&CK), the defined \acp{TTP} specific to the ICS environment primarily address the behaviors associated with the manipulation of Windows-based engineering workstations, safety controllers, and the associated network communications between them. 
The following sections will review the adversary activity associated with these three areas based on the previously defined criteria.
        
\textbf{TriStation Workstation (Table 1):} The Triton malware initially targeted the engineering workstation and associated TriStation software running on that device. 
The malware included a Python program that was compiled into an EXE, but which included a file name that masquerades as the authentic TriStation executable [T0849]. 
The malware then used the same \ac{API} that TriStation uses to communicate with and manage the Triconex devices to access those devices [T0834]. 
Because the workstation is a traditional Windows-based platform, both of these techniques could be determined by reviewing files matching the hash of the EXE file, leveraging the \ac{STIX} File: Hash object. 
This hash value was provided in the threat intelligence reports, and the Windows-based functionality is well-documented. 
However, detecting the malicious \ac{API} uses likely depends on the runtime analysis of the specific system calls used by the TriStation workstations, which are not specified in any documentation, and there is no available \ac{STIX} object to represent these system calls.

        \begin{table}[H]
        \begin{center}
            \caption{Analysis of Triton Software TriStation Workstation Activity.\label{Table:1}}
            \begin{tabular}{|m{3em}|m{6em}|m{7em}|m{5em}|m{5em}|m{7em}|m{4em}|}
            \hline
            Tech. & 1-Data Source & 2-Artifact & 3-STIX & 4-Artifact Detailedness & 5-Proprietary & 6-Parser\\
            \hline\hline
            \textbf{T0849} & File-Metadata & File hash of exe & Full - \textit{File:hashes} & Actionable & Proprietary - Documented & NA - host\\ \cline{1-5}
            \textbf{T0853} & Command/ Process & File hash of exe & Full - \textit{File:hashes} & Actionable & & \\ \cline{1-6}
            \textbf{T0871} & Process - \ac{OS} API Exec & TriStation API calls & No & Actionable & Proprietary - Undocumented & \\
            \hline
            \end{tabular}
        \end{center}
        \end{table}

        \begin{flushleft}
        \textbf{TriStation Protocol (Table 2):} The TriStation protocol, which is a proprietary protocol without any available open-source parsers, was heavily used throughout the attack. 
        The adversarial techniques using this protocol include discovering target devices [T0846], detecting and changing their operating mode [T0858] [T0868], downloading programs to the targeted safety controller [T0843], executing the program [T0871], and extracting programs from the device [T0845]. 
        Developing detections for this activity requires the ability to (i) understand and parse the TriStation protocol ~\cite{tristation_protocol_reversing}, (ii) identify specific functions of the protocol, and (iii) represent function payloads (e.g., program downloads contents). 
        The \ac{STIX} Network Traffic observable could be used to represent this activity, specifically the \textit{src/dst\_payload\_ref objects}, which reference specific strings within the protocol payloads.
        However, using the \textit{src/dst\_payload\_ref} \ac{STIX} objects for the TriStation protocols techniques, any detection utilizing these objects is likely to be fragile.  
        \end{flushleft}
        
        \begin{table}[H]
        \begin{center}
            \caption{Analysis of Triton Software TriStation Protocol Activity.\label{Table:2}}
            \begin{tabular}{|m{3em}|m{6em}|m{7em}|m{6em}|m{5em}|m{7em}|m{4em}|}
                \hline
                Tech. & 1-Data Source & 2-Artifact & 3-STIX & 4-Artifact Detailedness & 5-Proprietary & 6-Parser \\ \hline\hline
                \textbf{T0846} & Network Traffic Content & Parsing of TriStation Protocol – Device discovery & Partial – Network Traffic src/ dst\_payload\_ ref & Described & Proprietary - Undocumented & No \\ \cline{1-1} \cline{3-3}
                \textbf{T0858} &  & - Operating mode changes &  &  &  &  \\ \cline{1-1} \cline{3-3}
                \textbf{T0868} &  & - Operating mode status &  &  &  &  \\ \cline{1-1} \cline{3-3}
                \textbf{T0843} &  & - Program downloads/appends &  &  &  &  \\ \cline{1-1} \cline{3-3}
                \textbf{T0845} &  & - Program uploads &  &  &  &  \\ \cline{1-1} \cline{3-3} \cline{5-7} 
                \textbf{T0885} &  & Custom malware C2 payload &  & Actionable & NA & NA \\ \hline
            \end{tabular}
        \end{center}
        \end{table}

        \begin{flushleft}
        \textbf{Triconex (Table 3):} Triton’s targeting of the Triconex safety controller required multiple adversary techniques used to manipulate devices that execute custom logic. 
        Once the malicious program was downloaded to the device, it was executed on the device [T0821]. 
        Then, the malware gained excessive privileges by executing a vulnerable system call on the device [T0890], and manipulated the running firmware of the device [T0857] to maintain an implant. 
        Next, it enabled \ac{C2} [T0869] by manipulating a protocol handler to link to malicious code injected into the device [T0874]. 
        Detecting this activity requires the ability to extract and audit programs on the device, such as through program upload functions or integrity checks.
        This could potentially be detected by the Process: Hash \ac{STIX} object, while many devices produce CRC’s or checksums of deployed programs, these could be easily converted to a hash to support this object type. 
        These \acp{CRC} within the Triconex devices are a proprietary but documented function~\cite{triconex_safety_evaluation}. 
        Most of these other Triconex behaviors are difficult to detect because they manipulate the device’s run-time memory, which depends on the undocumented and proprietary operation of the device. 
        Detecting the exploitation of a vulnerable system call typically requires a runtime agent running on the device, while detecting the runtime manipulation in the underlying firmware requires mechanisms that perform integrity checks of a device’s \ac{OS} and runtime environments. 
        Therefore, these specific techniques likely cannot be detected through traditional detection platforms and \ac{STIX} based information sharing.
        \end{flushleft}
        
        \begin{table}[H]
        \begin{center}
            \caption{Analysis of Triton Software Triconex Activity.
            \label{Table:3}}
            \begin{tabular}{|m{3em}|m{6em}|m{7em}|m{5em}|m{5em}|m{7em}|m{4em}|}
                \hline
                Tech. & 1-Data Source & 2-Artifact & 3-STIX & 4-Artifact Detailedness & 5-Proprietary & 6-Parser \\ \hline
                \textbf{T0821} & Asset Software & PLC program CRC & Partial – Process: Hash & Actionable & Proprietary - Documented & NA - host \\ \cline{1-6}
                \textbf{T0890} & Process - OS API Exec & OS API calls & No & Described & Proprietary - Undocumented &  \\ \cline{1-1}
                \textbf{T0834} &  &  &  &  &  &  \\ \cline{1-4}
                \textbf{T0857} & Firmware & FW/Memory Contents & Partial – Process: Hash &  &  &  \\ \cline{1-1}
                \textbf{T0874} &  &  &  &  &  &  \\ \hline
            \end{tabular}
        \end{center}
        \end{table}

    \subsubsection{Stuxnet} The Stuxnet malware targeted Siemens \acp{PLC} used to support uranium enrichment cascades. 
    The malware targeted both the \acp{PLC}, by manipulating the application program deployed on the device, and the engineering software deployed on the workstation (WinCC). 
    Therefore, the associated \acp{TTP} then heavily focuses on the manipulation of the \ac{PLC}, the software on the engineering workstation, and their associated network communications. 
    Tables 4-6 enumerate each adversarial technique defined within ATT\&CK for ICS associated with the Stuxnet malware.

        \begin{flushleft}
        \textbf{SIMATIC WinCC Workstation (Table 4):} The malware was initially deployed to engineering workstations using the SIMATIC WinCC software using either hard-coded credentials [T0891] or by copying itself to connected USB drives [T0847]. 
        The hard-coded credential used could have been represented by the \ac{STIX} \textit{User Account: Credential} observable to enable sharing this information. 
        The malware also propagates by modifying the content of SIMATIC Step 7\footnote{https://www.siemens.com/global/en/products/automation/industry-software/automation-software/tia-portal/software/step7-tia-portal.html} project files found on the device by searching for files with the appropriate Step 7 format [T0873]. 
        The modified project files could be detected by sharing the malicious contents or by representing the overall hashes of the activity, either through the \ac{STIX} \textit{File: String} or \textit{File: Hash} observable.  
    
        Then, the malware installs a malicious \ac{DLL} which masquerades as an authentic file [T0849], the hash of this file could be represented by the \textit{File: Hash} \ac{STIX} observable. 
        This DLL further injects any project files loaded [T0863]. 
        The malicious DLL identifies targets by reviewing the contents of incoming data blocks to identify specific devices [T0888]. 
        The \ac{DLL} then downloads code to the \ac{PLC} [T0843] and manipulates data sent between the PLC and workstation [T0851]. 
        Since these activities all require understanding the execution of \ac{DLL}, detecting this activity would require an understanding of specific function calls both within the \ac{DLL} and \ac{OS} system calls. 
        However, as previously stated, \ac{STIX} currently lacks support for this.
        \end{flushleft}
    
        \begin{table}[H]
        \begin{center}
            \caption{Analysis of Stuxnet WinCC Workstation Activity.\label{Table:4}}
            \begin{tabular}{|m{3em}|m{6em}|m{7em}|m{5em}|m{5em}|m{7em}|m{4em}|}
                \hline
                Tech. & 1-Data Source & 2-Artifact & 3-STIX & 4-Artifact Detailedness & 5-Proprietary & 6-Parser \\ \hline
                \textbf{T0891} & Logon Session - Metadata & SQL Server password & Full – User Acct: Cred & Actionable & Proprietary - Documented & NA - host \\ \hline
                \textbf{T0847} & File - Metadata & Malicious file names & Full – File: Name & Actionable & Proprietary - Documented & NA - host \\ \hline
                \textbf{T0873} & File - Metadata & PLC Project Files & Full – File: String/Hash & Described & Proprietary - Documented & NA - host \\ \hline
                \textbf{T0849} & File - Metadata & DLL hash & Full – File: String/Hash & Actionable & Proprietary - Documented & NA - host \\ \hline
                \textbf{T0863} & Command Execution / Process Creation & Command name & Full - File: Name & Described & Proprietary - Documented & NA - host \\ \hline
                \textbf{T0874} & Process – OS API Exec & OS API calls & No & Mentioned & Proprietary - Documented & NA - host \\ \cline{1-1}
                \textbf{T0888} &  &  &  &  &  &  \\ \cline{1-1}
                \textbf{T0851} &  &  &  &  &  &  \\ \hline
            \end{tabular}
        \end{center}
        \end{table}

        \begin{flushleft}
        \textbf{Network Activity (Table 5):} Stuxnet used a wide array of network activity to both propagate malware across devices and also to manipulate the control of the operational process. 
        The malware utilized various SQL server-stored procedures to transfer files and execute malware [T0886], including transferring the Stuxnet \ac{DLL} to other systems [T0867] and executing the malware [T0866]. 
        This would require monitoring the SQL Server protocol for the execution of those specific stored procedures. 
        This could be potentially shared by using the \ac{STIX} \textit{Network Traffic: src/dst\_payload\_ref} observable; however, it is unclear whether the SQL protocols could be effectively represented by payload strings. 
        Then, the malware attempts to establish a \ac{C2} connection to external servers [T0885] over port 80, which could have been shared using the \ac{STIX} \textit{Network Traffic: dst\_port} observable.

        The Siemens S7Comm protocol was used to perform program downloads from the WinCC workstation to the \acp{PLC} [T0843]. Once the malicious code was deployed to the \ac{PLC}, it then sends Profibus messages to frequency converter drives [T0836]. 
        While Profibus is an \ac{IEC} standard, S7Comm (Siemens) is a proprietary protocol, though both have publicly available parsers. 
        These activities would also likely need to be shared using the \ac{STIX} \textit{Network Traffic: src/dst\_payload\_ref} observable, though it is unclear whether these behaviors could be effectively shared using it.
        \end{flushleft}
        
        \begin{table}[H]
        \begin{center}
            \caption{Analysis of Stuxnet Network Activity.\label{Table:5}}
            \begin{tabular}{|m{3em}|m{6em}|m{7em}|m{6em}|m{5em}|m{7em}|m{4em}|}
                \hline
                Tech. & 1-Data Source & 2-Artifact & 3-STIX & 4-Artifact Detailedness & 5-Proprietary & 6-Parser \\ \hline
                \textbf{T0866} & Network Traffic Content & SQL Strings/ Protocol Parser & Partial – Network Traffic: src/ dst\_payload\_ ref & Actionable & Proprietary - Documented & Yes \\ \cline{1-1}
                \textbf{T0867} &  &  &  &  &  &  \\ \hline
                \textbf{T0885} & Network Traffic Connection & Connection to Port 80 & Full – Network Traffic: dst \_port & Mentioned & Proprietary - Documented & NA \\ \hline
                \textbf{T0843} & Network Traffic Content & Parsing of S7Comm Protocol Functions: Program Downloads/Appends & Partial – Network Traffic: src/ dst\_payload\_ ref & Missing & Proprietary - Undocumented & Yes \\ \hline
                \textbf{T0836} & Network Traffic Content & Profibus Network Traffic & Partial – Network Traffic: src/ dst\_payload\_ ref & Described & Open / Standard & No \\ \hline
            \end{tabular}
        \end{center}
        \end{table}

        \begin{flushleft}
        \textbf{Siemens PLCs (Table 6):} The malware deploys malicious programs by manipulating controller tasking (e.g., OB1, OB35) [T0851][T0821] on the \acp{PLC}. 
        The manipulation of controller tasking and programs can be detected by either performing program uploads or performing integrity checks. 
        These could be partially represented by \textit{File: Hash} \ac{STIX} observables, since many \ac{PLC} application programs are frequently represented by other message digest mechanisms, such as \acp{PLC} or checksums. 
        This malicious program then performs a number of additional techniques on the \acp{PLC}, including sniffing Profibus message contents [T0842] and monitoring the state of the operational process [T0801] and calls an existing system function block used by the device [T0834]. 
        The program collects data from the I/O image [T0877] and then intercept/overwrite peripheral output to prevent detection [T0835]. 
        These could either be detected through some sort of run-time analysis, by verifying the specific \ac{OS} \ac{API} calls made, or by extracting and performing an analysis of the malicious program performed on the device. 
        The former would depend on the device’s ability to perform run-time monitoring of device functions; this capability is typically not supported by \acp{PLC}. Since most PLCs do support application program uploads, the specific \ac{PLC} \ac{API} calls can be analyzed to observe this behavior. 
        Unfortunately, this presents a clear gap in \ac{STIX}, as there is currently no way to reference specific \ac{PLC} application logic functions.
        \end{flushleft}
        
        \begin{table}[H]
        \begin{center}
            \caption{Analysis of Stuxnet PLC Activity.\label{Table:6}}
            \begin{tabular}{|m{3em}|m{8em}|m{5em}|m{5em}|m{5em}|m{7em}|m{4em}|}
                \hline
                Tech. & 1-Data Source & 2-Artifact & 3-STIX & 4-Artifact Detailedness & 5-Proprietary & 6-Parser \\ \hline
                \textbf{T0821} & Asset – Software & PLC code & Partial – File: Hash & Described & Proprietary - Documented & NA - Host \\ \hline
                \textbf{T0842} & Asset – Software / Process – OS API Exec & PLC OS API calls & No & Described & Proprietary - Documented & NA - Host \\ \cline{1-3}
                \textbf{T0801} & Asset – Software / Process – OS API Exec & PLC OS API calls &  &  &  &  \\ \cline{1-3} \cline{5-5}
                \textbf{T0869} & Asset – Software / Process – OS API Exec & PLC OS API calls &  & Actionable &  &  \\ \cline{1-3} \cline{5-5}
                \textbf{T0834} & Asset – Software / Process – OS API Exec & PLC OS API calls &  & Described &  &  \\ \cline{1-3} \cline{5-5}
                \textbf{T0877} & Asset – Software / Process – OS API Exec & PLC OS API calls &  & Actionable &  &  \\ \cline{1-3} \cline{5-5}
                \textbf{T0835} & Asset – Software / Process – OS API Exec & PLC OS API calls &  & Described &  &  \\ \hline
            \end{tabular}
        \end{center}
        \end{table}

    \subsubsection{Industroyer} The Industroyer or CrashOverride malware was used as part of an attack that occurred in December 2016, targeting Ukraine's power grid and causing a blackout in the northern part of the capital, Kyiv. 
    It was capable of operating a substation’s switches and circuit breakers, utilizing several common industrial communication protocols, including IEC 60870-5-101, IEC 60870-5-104, IEC 61850, and OPC DA. 
    Additionally, this malware features a built-in "wiper" module that can erase the configuration files for the equipment controlling the circuit breakers, making recovery more challenging.

        \textbf{Windows Activity (Table 7):} The malware launcher is initially executed as a .dll through a command line parameter [T0807], and multiple other modules were also executed as .dlls. 
        This could be represented as a \ac{STIX} File Name object, while the initial launcher .dll name is not included in the report, the associated wiper .dlls are provided. 
        The malware then enumerates systems call on the system, which requires executing Windows system calls [T0840]. 
        Furthermore, the malware connects to COM ports on the Windows system, which involves executing Windows system calls. However, neither of these system-call-focused artifacts can be represented by \ac{STIX} observables [T0803] [T0804] [T0805]. 
        Later, the malware executes a wiper that disables system services and wipes files on the device. 
        While the specific system calls used to perform this are not defined, there are likely a limited set of Windows calls that enable this. 
        Furthermore, \ac{STIX} lacks an observable to represent system calls used to understand runtime malware functions.

        \begin{table}[H]
        \begin{center}
            \caption{Analysis of Industroyer Windows Activity.\label{Table:7}}
            \begin{tabular}{|m{3em}|m{6em}|m{8em}|m{5em}|m{5em}|m{7em}|m{4em}|}
                \hline
                Tech. & 1-Data Source & 2-Artifact & 3-STIX & 4-Artifact Detailedness & 5-Proprietary & 6-Parser \\ \hline
                \textbf{T0807} & Command Execution & CMD line parameter w/ DLL & Full – File: Name & Actionable & Proprietary - Documented & NA - Host \\ \hline
                \textbf{T0881} & Service - Metadata / Process – OS API Exec & OS API call associated with service stop & No & Actionable & Proprietary - Documented & NA - Host \\ \hline
                \textbf{T0809} & Process – OS API Exec & OS API calls associated with wiper functions & No & Actionable & Proprietary - Documented & NA - Host \\ \hline
                \textbf{T0840} & Process – OS API Exec & OS API calls associated with network enumeration & No & Actionable & Proprietary - Documented & NA - Host \\ \hline
                \textbf{T0803} & Process – OS API Exec & System Calls for COM Ports & No & Mentioned & Proprietary - Documented & NA - Host \\ \cline{1-1}
                \textbf{T0804} &  &  &  &  &  &  \\ \cline{1-1}
                \textbf{T0805} &  &  &  &  &  &  \\ \hline
            \end{tabular}
        \end{center}
        \end{table}

        \textbf{Network Activity (Table 8):} The first network-focused activity is the establishment of a proxy through port 3128 [T0884], which can be represented by the \ac{STIX} \textit{Network Traffic src/dest\_port} observable. 
        Then, the malware has a broad set of behaviors that utilize a set of automation protocols to interact with substation devices, including supporting remote discovery, collection, monitoring, and sending unauthorized command messages.

        At least two protocols, OPC and 61850 leveraged discover functions within the protocols to identify devices supporting those protocols [T0846]. 
        Then, several protocol functions were utilized to collect and monitor the status of devices using the 61850, OPC, and IEC 104 protocols [T0888][T0801]. 
        The unauthorized command messages were used to manipulate the operation of devices, specifically to open/close breakers [T0855][T0806]. 
        While none of these protocols have specific representation in \ac{STIX}, they could be partially represented by the \textit{Net Traffic src/dst\_payload\_ref} object. 
        Fortunately, these are all open or standards-based protocols, and all have available parsers.

        \begin{table}[H]
        \begin{center}
            \caption{Analysis of Industroyer Network Activity.\label{Table:8}}
            \begin{tabular}{|m{3em}|m{6em}|m{8em}|m{6em}|m{5em}|m{5em}|m{4em}|}
                \hline
                Tech. & 1-Data Source & 2-Artifact & 3-STIX & 4-Artifact Detailedness & 5-Proprietary & 6-Parser \\ \hline
                \textbf{T0884} & Network Traffic Connection & Connection to Port 3128 & Full – Network Traffic: src/dest\_port & Described & Open / Standard & NA \\ \hline
                \textbf{T0846} & Network Traffic Content & 61850 discovery functions & Partial Network Traffic: src/ dst\_payload\_ ref & Described & Open / Standard & Yes \\ \hline
                \textbf{T0846} & Network Traffic Content & OPC discovery functions & Partial Network Traffic: src/ dst\_payload\_ ref & Actionable & Open / Standard & Yes \\ \cline{1-1} \cline{3-3}
                \textbf{T0801} &  & OPC/61850 monitor functions &  &  &  &  \\ \cline{1-1} \cline{3-3}
                \textbf{T0888} &  & 61850 collection functions &  &  &  &  \\ \cline{1-1} \cline{3-3}
                \textbf{T0888} &  & OPC collection functions &  &  &  &  \\ \cline{1-1} \cline{3-3}
                \textbf{T0888} &  & 104 collection functions &  &  &  &  \\ \cline{1-1} \cline{3-3}
                \textbf{T0855} &  & 104 command functions &  &  &  &  \\ \cline{1-1} \cline{3-3}
                \textbf{T0806} &  & 104 command functions &  &  &  &  \\ \hline
            \end{tabular}
        \end{center}
        \end{table}
        
        \begin{flushleft}
        \textbf{SIPROTEC/Relay Activity (Table 9):} Industroyer also has the capability to use the CVE-2015-5374 vulnerability to make a SIPROTEC relay unresponsive. The malware sends a SIPROTECH message to a relay, causing the device to enter an unresponsive state, requiring the device to be rebooted [T0814][T0816]. 
        This message is an 18-byte packet to a proprietary UDP port 50,000 on the device.
        This 18-byte string is provided in the threat intelligence and could be represented by the \ac{STIX} payload\_ref object; however, it is unclear whether this is a reliable or precise indicator, as it is a proprietary protocol. 
        On the device side, the message initiates an incomplete firmware update, which is not completed, leaving the device unresponsive [T0800]. 
        This is potentially available in device application logs; however, there is no detail about this activity from the device side, and no existing \ac{STIX} observable to support log events. 
        \end{flushleft}

        \begin{table}[H]
        \begin{center}
            \caption{Analysis of Industroyer Network Activity.\label{Table:9}}
            \begin{tabular}{|m{3em}|m{6em}|m{7em}|m{6em}|m{5em}|m{7em}|m{4em}|}
                \hline
                Tech. & 1-Data Source & 2-Artifact & 3-STIX & 4-Artifact Detailedness & 5-Proprietary & 6-Parser \\ \hline
                \textbf{T0814} & Network Traffic Content & Strings/ SIPROTECH protocol parser \& mgmt functions & Partial – Network Traffic: src/ dst\_payload\_ ref & Actionable & Proprietary - Undocumented & No \\ \cline{1-1}
                \textbf{T0816} &  &  &  &  &  &  \\ \hline
                \textbf{T0800} & Application Log Content & Device Log Events & No & Missing & Proprietary - Undocumented & NA - host \\ \hline 
            \end{tabular}
        \end{center}
        \end{table}
        
\subsection{Thorough Analysis of Threat Information Sharing at ATT\&CK for ICS}
\subsubsection{Methodology}
    To provide a comprehensive assessment beyond the three detailed case studies, we conducted a systematic analysis of all 196 procedure examples across 79 techniques documented in ATT\&CK for ICS. 
    Our methodology consists of four phases:
    \begin{enumerate}
        \item \textbf{\textit{Data Acquisition and Preparation}} - All procedure descriptions were retrieved from the public MITRE ATT\&CK \ac{STIX} data for ICS\footnote{https://github.com/mitre-attack/attack-stix-data/blob/master/ics-attack/ics-attack.json} and normalized into plain-text records. Each record retained its corresponding technique identifier, and malware family linkage.
    
        \item \textbf{\textit{Automated Observable Extraction}} - Each procedure description was processed by a \ac{LLM} (GPT o3-mini) using a structured prompt that explicitly defined the task, decision rules, and expected JSON output format. 
        The model was instructed to identify every observable mentioned in the text, ranging from file names and commands to ICS-specific protocol elements, and to classify each one according to the pre-defined attributes.
        
        \item \textbf{\textit{Observable Categorization}} - Every extracted observable was represented as a structured schema record containing the key criteria we defined before.
        \item \textbf{\textit{Post-processing and Quality Control}} - According to the \ac{STIX} 2.1 specification, the software Cyber-Observable Object is defined as "high-level properties associated with software, including software products".\footnote{https://docs.oasis-open.org/cti/stix/v2.1/csprd01/stix-v2.1-csprd01.html\#\_Toc16070740} 
        This equivocal definition can mislead an LLM into labelling malware identifiers as observables. 
        Yet, \ac{STIX} classifies malware itself as a distinct Domain Object\footnote{https://docs.oasis-open.org/cti/stix/v2.1/csprd01/stix-v2.1-csprd01.html\#\_Toc16070566} rather than as an observable artifact. 
        Because \acp{SCO} are intended to capture low-level forensic evidence, while Domain Objects (e.g., malware encodes higher-order threat-intelligence concepts), keeping malware entries in the observable layer would blur these semantic boundaries.\footnote{https://docs.oasis-open.org/cti/stix/v2.1/csprd01/stix-v2.1-csprd01.html\#\_Toc16070567}
        To preserve the intended separation of concerns, we therefore filtered out every item that the model had incorrectly flagged as malware-related.
        The resulting dataset was manually reviewed to eliminate duplicates, to remove entities incorrectly labeled as Observables, and to correct misclassifications.
        Overall, out of 361 observables, 103 observables' data were manually modified:
        \begin{itemize}
        \item 34 observables specifications were modified to be more ICS specific domain
        \item 27 observables were manually classified as \textit{Actionable} instead of \textit{Described}
        \end{itemize}    
    \end{enumerate}

    \begin{mdframed}[style=promptbox]
        Extracted Observable example:
        \begin{lstlisting}[language=json,basicstyle=\tiny\ttfamily,breaklines=true,frame=single]
        {
            'observable_value': 'CSW', 
            'artifact_details': 'Described', 
            'data_source': 'ICS historian', 
            'classification': 'ICS Data Tag', 
            'STIX_supported': 'No', 
            'proprietary_artifact': 'Open/Standard Technology', 
            'parser': 'libiec61850', 
            'notes': 'Logical-node data attribute indicating circuit-breaker / switch control capability.', 
            'description': "The [Industroyer](https://attack.mitre.org/software/S0604) IEC 61850 component sends the domain-specific MMSgetNameList request to determine what logical nodes the device supports. It then searches the logical nodes for the CSW value, which indicates the device performs a circuit breaker or switch control function.(Citation: ESET Industroyer)\n\n[Industroyer](https://attack.mitre.org/software/S0604)'s OPC DA module also uses IOPCBrowseServerAddressSpace to look for items with the following strings: ctlSelOn, ctlOperOn, ctlSelOff, ctlOperOff, Pos and stVal.(Citation: ESET Industroyer)\n\n[Industroyer](https://attack.mitre.org/software/S0604) IEC 60870-5-104 module includes a range mode to discover Information Object Addresses (IOAs) by enumerating through each.(Citation: ESET Industroyer)",
            'technique_num': 'T0888', 
            'description_id': 'relationship--62e818b8-38e6-42ff-9424-9a327332eb2a',
            'related_malware': 'Industroyer'
        }
        \end{lstlisting}
        
    \end{mdframed}

        The systematic analysis revealed patterns that directly address our research questions. The following sections present the empirical findings, organized by: (i) \ac{STIX} representation gaps; (ii) actionability of extracted observables; and (iii) dependency on proprietary technologies.

\subsubsection{Observation and Analysis}
\label{subsec:Observation and Analysis}
\paragraph{STIX GAP Taxonomy}

Out of 361 extracted observables, over half of the observables (191, around 53\%) have only partial support in \ac{STIX} 2.1, meaning they can only be approximated via generic or
custom objects. 
About 19\% (69 observables) have no support at all in \ac{STIX} (i.e., no relevant cyber observable exists), while only around 28\% (101 observables) are fully supported by an existing \ac{STIX} object type.
This indicates a significant coverage gap for many ICS-specific artifact types.
\begin{enumerate}
    \item \textbf{Fully Supported (101)} - These are mostly standard \ac{IT} indicators (e.g., files, network addresses, domain names) which map cleanly to \ac{STIX} \acp{SCO} like \textit{File}, \textit{IPv4-Addr}, \textit{Domain-Name}, etc.
    Few ICS-specific items fall in this category, underscoring that \ac{STIX} 2.1’s built-in types were not designed with \ac{ICS} nuances in mind.
    
    \item \textbf{Partially Supported (191)} - The majority of \ac{ICS} observables fall here. 
    They often correspond to \ac{ICS} domain concepts that lack a first-class \ac{STIX} object, but can be accommodated within a generic object or a custom extension.
    For example: \textit{MMSgetNameList} (an \ac{ICS} protocol command) is labeled \textit{Partial: network-traffic}, meaning it can be captured as network traffic data, but \ac{STIX} has no explicit "ICS command" type.
    Many \ac{PLC} programming functions (e.g., Siemens S7 functions like \textit{SFC1} or block names like \textit{OB1}) were marked \textit{Partial: artifact}, indicating they could only be represented as generic file/artifact content.
    
    \item \textbf{Not Supported (69)} -  These are observables with no meaningful \ac{STIX} representation. 
    They include highly ICS-specific elements for which even an approximation is lacking.
    For example: \textit{ctlSelOn} – an \ac{ICS} data tag/field from a control system historian – is marked "No support" because \ac{STIX} has no concept of a tag or point name in a \ac{PLC}/\ac{SCADA} database.
    Likewise, \ac{ICS} memory addresses and register values (e.g., \acp{IOA} in IEC-104) are also unsupported.
    In general, ICS data tags, device configurations, and field values (e.g., status flags, function codes) show up repeatedly among the "No support" entries.
    This trend highlights that industrial control system observables are a major gap area in \ac{STIX} 2.1, the schema covers \ac{IT}-centric artifacts well, but things like control logic variables, \ac{PLC} memory blocks, or proprietary protocol fields fall outside its current scope.
\end{enumerate}

\paragraph{Trends in Unsupported Types.}

The unsupported entries tend to cluster in specific classifications. 
Many \textit{ICS Data Tag artifacts} (control system historian points, device tags) had no \ac{STIX} support.
Even some general \ac{IT} items like specific software tool names or report identifiers were marked unsupported when they do not correspond to \ac{STIX}'s defined objects (e.g., an internal report ID or a code function name has no direct \ac{STIX} analog).
In contrast, common network or host indicators, such as files, IP addresses, and user accounts, are typically fully covered.
This suggests that \ac{STIX} 2.1's coverage gap is most pronounced for specialized \ac{ICS} domain artifacts, such as data related to industrial devices, proprietary protocols, and control logic, whereas typical enterprise observables are well-supported.

\paragraph{Artifact Detail Level Distribution}

    \begin{figure}[H]
                \centering 
                \includegraphics[width=0.9\textwidth]{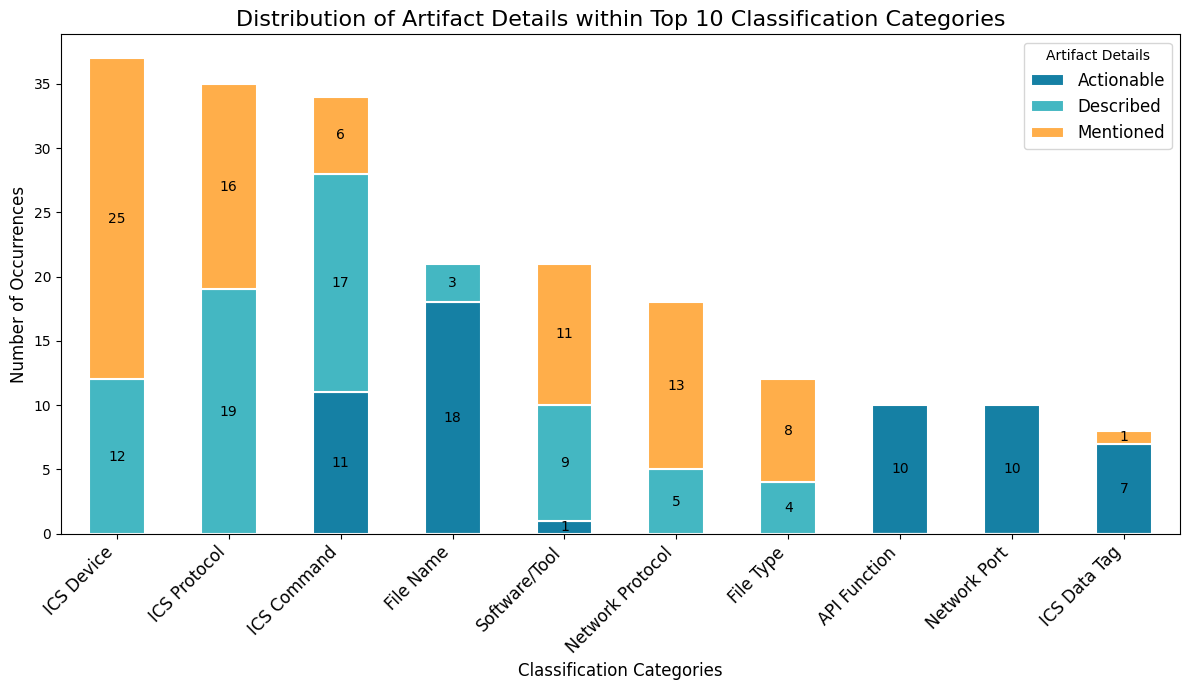}
                \caption{Distribution of Artifact Details within Top 10 Classification Categories ("Actionable" in blue, "Described" in teal, "Mentioned" in orange).}
                \label{fig:distribution_top_10_categories}
    \end{figure}

    The observables were categorized into over 100 classification types, but a few dominate (see Figure 2). It is worth noting that an ICS data artifact class, ICS Data Tag (8), also makes the top 10, indicating that specific tag names/variables from control system databases were noted multiple times despite \ac{STIX} gaps. 
    In general, the classification distribution highlights that \ac{ICS}/\ac{OT} entities (devices, protocols, and commands) were prominent, alongside a mix of traditional \ac{IT} indicators (files and network configurations) present in the threat reports.

    \begin{figure}[H]
                \centering 
                \includegraphics[width=0.8\textwidth]{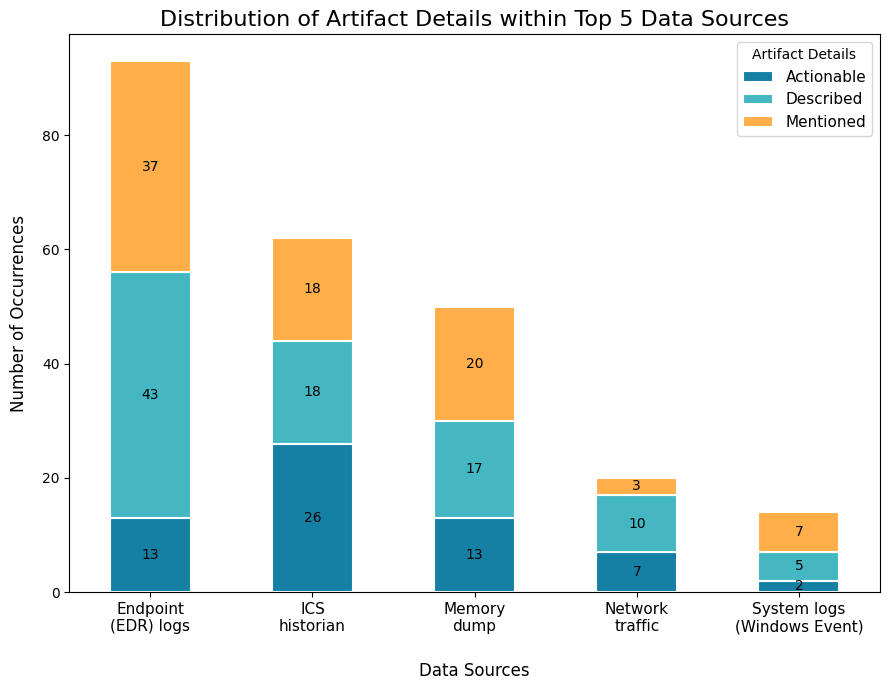}
                \caption{Distribution of Artifact Details within Top 5 Data Sources("Actionable" in blue, "Described" in teal, "Mentioned" in orange).}
                \label{fig:distribution_top_5_datasources}
    \end{figure}

    Each observable was also tagged with a data source indicating where that artifact could be collected. 
    The most common source by far is network traffic (93 observables). 
    This aligns with the many \ac{ICS} protocol messages and network indicators extracted (since those would be seen in network captures or \acp{PCAP}).
    Notably, ICS historian logs (50) are the third-largest category; many of the ICS-specific observables, such as tag values and device status changes, are recorded in historian databases or engineering logs, emphasizing the importance of historian data in detecting \ac{ICS} attacks.
    It is worth highlighting that ICS-specific telemetry (Historian, \ac{PLC} ladder logic captures) together make up a significant chunk, demonstrating that to fully monitor these observables, one must pull from \acp{ICS} and not just traditional \ac{IT} logs.

\paragraph{Actionable Observables Lacking \ac{STIX} Support}

    One important observation is that several observables deemed "Actionable" (high-value \acp{IOC}) are not fully supported in \ac{STIX} 2.1, which could hinder information sharing. 
    In our results, 48 actionable observables (over half of all actionable ones) were only partial or not supported in \ac{STIX}. 
    These include ICS-specific items, such as \textit{ctlSelOn}, which is a precise, unique tag name that an attacker can manipulate (actionable for detection in \ac{ICS} logs); yet, \ac{STIX} has no way to represent it except as a generic artifact or custom field. 
    The same goes for certain \ac{PLC} function codes or protocol commands that are very specific (and thus actionable within that context) but have no \ac{STIX} presentation.

    The key insight is that \textbf{being "Actionable" does not guarantee \ac{STIX} support.} 
    There is a misalignment where \ac{CTI} producers can identify a highly specific \ac{ICS} indicator, but when codifying it in \ac{STIX} format to share with others, they struggle due to the schema's limitations.
    This can complicate the intelligence sharing process for \ac{ICS} threats; either the observable has to be dropped (if there's no representation) or shared in an unstructured way (in notes or description fields), reducing its utility. 
    It reinforces the earlier point that extending \ac{STIX} or developing custom \acp{SCO} for \ac{ICS} could significantly enhance our ability to share and leverage these crucial observables. 
    The fact that dozens of actionable \ac{ICS} indicators fall through the cracks in \ac{STIX} 2.1 is a call to action for the community to address these schema gaps for industrial security.

\section{Analysis of Vulnerability Information Sharing}
\label{sec:Analysis of Vulnerability Information Sharing}
    In addition to studying cyber threat activity, the analysis also reviews the sharing of vulnerability advisories associated with known exploitable \ac{ICS} vulnerabilities based on their inclusion in the \ac{CISA} \ac{KEV} Catalog. 
    This list was produced by reviewing the \ac{KEV} Catalog for \acp{CVE} identified since 2023, which are associated with \ac{CISA} \ac{ICS} Alerts. 
    Since these vulnerabilities have been known to be exploited by threat actors, they present a clear risk to asset owner environments, requiring some mitigating actions. 
    While within IT environments these vulnerabilities are typically patched, it is well documented asset owners struggle to quickly deploy patches~\cite{sans_ics_ot_cybersecurity}. 
    If an asset owner cannot quickly patch these vulnerabilities to mitigate their risk, they may need to deploy detection mechanisms to prevent their exploitation. 
    Therefore, vulnerability information is reviewed in accordance with the same requirements as defined in Section 2. 
    That is, evaluating whether the vendor provided some rule to support the detection of the vulnerability, or whether sufficient technical details are available about the product and associated vulnerability for the asset owner to generate their own detection.

    Table 10 documents the analysis of the 9 \ac{CISA} vulnerability advisories and the associated vendor-provided vulnerability advisories. 
    For each advisory, the \ac{CISA} \ac{ICS} Advisory number, the associated \acp{CVE}, vendor, and associated protocols (if applicable) are defined. 
    Each advisory was reviewed for the following criteria.
    \begin{enumerate}
        \item \textit{Rules/Analytics Available (1):} This identifies whether the advisory provides any information about detection rules (e.g., a Snort rule) or analytics that can be deployed to detect the advisories.

        \item \textit{Availability of Technical Details (2):} This criterion assesses whether the report or advisory includes sufficient technical information such that a detection rule or analytic could be developed based on that information. 
        While many vulnerability advisories provide high-level discussion about the vulnerability and mappings to known \acp{CWE}, they also do not provide the technical details required to develop rules or analytics to detect exploitation.
        If technical information is available but insufficient to inform a detection, it is labeled as 'Partial.'

        \item \textit{Proprietary (3), and Parsers (4):} For vulnerabilities that are associated with a network protocol, this explores whether the protocol is open, proprietary, and documented, or proprietary and undocumented (3). Next, it reviews whether a parser exist for that protocol (4). If a parser exists but does not fully cover the protocol’s use or implementation, it is labeled as Partial.
    \end{enumerate}

    \begin{table}[H]
    \begin{center}
        \caption{Table 10: Review of 9 ICS vulnerabilities with known exploitation.\label{Table:10}}
        \begin{tabular}{|l|l|l|l|l|l|l|l|}
        \hline
        ICSA Alert & CVE & Vendor & Protocol & 1- Rules & 2 - Tech Detail & 3 - Proprietary & 4 - Parsers \\ \hline
        \textbf{23-355-01} & 2023-49897 & FXC & HTTP & No & No & Open - HTTP & Partial \\ \hline
        \textbf{23-355-02} & 2023-47565 & QNAP & HTTP & No & No & Open - HTTP & Partial \\ \hline
        \textbf{23-348-15} & \begin{tabular}[c]{@{}l@{}}2023-4911 \\ 2023-44487\end{tabular} & Siemens & NA - Host & No & Partial & N/A & N/A \\ \hline
        \textbf{23-320-11} & 2023-6448 & Unitronics & PCOM & No & No & \begin{tabular}[c]{@{}l@{}}Proprietary - \\ Undocumented\end{tabular} & No \\ \hline
        \textbf{23-297-01} & \begin{tabular}[c]{@{}l@{}}2023-20273 \\ 2023-20198\end{tabular} & \begin{tabular}[c]{@{}l@{}}Rockwell \\ Automation\end{tabular} & HTTP & \begin{tabular}[c]{@{}l@{}}Yes - \\ External\end{tabular} & Partial & Open - HTTP & Partial \\ \hline
        \textbf{23-264-05} & 2020-16017 & \begin{tabular}[c]{@{}l@{}}Rockwell \\ Automation\end{tabular} & HTTP & No & Partial & Open - HTTP & Partial \\ \hline
        \textbf{23-075-01} & 2021-4034 & Siemens & NA - Host & No & Yes & N/A & N/A \\ \hline
        \begin{tabular}[c]{@{}l@{}}\textbf{23-193-01} \\ Documented  \\ exploit but  \\ not in KEV\end{tabular} & \begin{tabular}[c]{@{}l@{}}2023-3595 \\ 2023-3596\end{tabular} & \begin{tabular}[c]{@{}l@{}}Rockwell \\ Automation\end{tabular} & CIP & Yes & Yes & \begin{tabular}[c]{@{}l@{}}Proprietary - \\ Documented\end{tabular} & Yes \\ \hline
        \end{tabular}
    \end{center}
    \end{table}

\subsection{Key Outcomes}

    The outcome of this analysis highlights that current vulnerability advisories lack sufficient technical details for asset owners to create their own detections. 
    Only two out of the nine vendor advisories reviewed include a detection rule or analytic for the vulnerabilities. 
    Furthermore, vulnerability advisories often lack the necessary technical details to support the development of effective detections. 
    Four provided no technical details, excluding the two vulnerabilities that included detection rules, while three provided partial technical discussions, which were insufficient for developing a detection. Further, one protocol was fully proprietary and undocumented. 
    At the same time, five were based on HTTP, an open protocol, but can have highly diverse specific customization, making parsing difficult without details of the specific application functions.
    While parsers for HTTP exist, they would still need to be heavily tailored towards the particular device’s application.
    This analysis highlights that detection rules are not consistently provided and critical information about device vulnerabilities or associated protocols not provided with sufficient detail to enable the development of detections. 
    Based on this analysis, it’s unclear how asset owners are expected to protect themselves from the reported vulnerabilities.

\section{Challenges in Threat Information Sharing}
\label{sec:Challenges in Threat Information Sharing}

    The prior sections provide an evidence-driven review of both available threat intelligence data (through ATT\&CK for ICS), and recent vulnerabilities advisories defined by vendors and \ac{CISA}.
    This analysis identifies four key challenges that should be addressed to ensure threat information can be more shared more effectively to ensure asset owners can effectively detect these threats.
    \begin{enumerate}
    
        \item \textbf{A lack of defined technical artifacts to support sharing of indicators of OT threats through \ac{STIX}.} This analysis demonstrated that while existing \ac{STIX} Observables would be sufficient for detecting a large majority of the TTPs identified in the previous analysis, there remain multiple TTPs that lack sufficiently defined technical artifacts. Therefore, expanding \ac{STIX} observables is necessary to support the sharing of all known adversarial techniques associated with \ac{ICS} incidents. Key examples include:
        \begin{itemize}
            \item \textit{PLC Code/Logic:} Numerous techniques include the deployment of malicious logic on a \ac{PLC}. 
            While many \acp{PLC} leverage vendor-specific tailoring of their logic, many devices share some fundamental capabilities and languages, especially as they align with the IEC 61131-3 standard for developing programs. 
            These programs are typically compiled before deployment to the controller. 
            However, the source code is typically transferred alongside the binaries to support diagnostics of the deployed program. 
            Therefore, mechanisms to support understanding program segments on the controller, both through source and compiled code, may be effective in detecting malicious logic.
            
            \item \textit{Integrity checks:} Further, many controllers also use integrity check mechanisms, typically based on \acp{CRC} or Checksums, to verify that programs have not changed ~\cite{plc_security_practices}. 
            On the one hand, security products can verify that a specific program has been downloaded to the device by validating its signatures. 
            On the other hand, \ac{CRC}/Checksum integrity checks are not sufficient against threat actors who intentionally modify the \ac{PLC} code ~\cite{schuett2014evaluation}. 
            There is a need for a wider deployment of secured integrity signatures ~\cite{banerjee2024designing}.

            \item \textit{ICS Protocols:} Currently, the \ac{STIX} standard supports prevalent \ac{IT} protocols, including HTTP and SMTP. However, each incident reviewed above has techniques that utilize ICS-specific protocols. 
            While these could potentially be parsed through more general \textit{NetworkTraffic} payload objects, the associated observables will still likely be ineffective without tailored objects. 
            Therefore, developing observable objects to support key \ac{ICS} protocols and fields would improve the effectiveness of sharing and detection of techniques targeting those protocols. 
            It may be possible to craft a general \ac{ICS} protocol object using typical functionality associated with \ac{ICS} protocols.
        \end{itemize}
        
        \item \textbf{The dependence on proprietary technologies and protocols without available parsers hinders the ability to develop detections.}
        While vendor technologies may require custom/proprietary protocols or the customization of standard protocols to address unique challenges faced when interfacing with devices, these protocols are increasingly targeted by threat actors, and therefore necessitate the ability to be monitored for threats by the asset owner. 
        The following table reviews the unique protocols identified in this analysis. 
        Of the nine overall protocols, at least three proprietary protocols were identified, and two had no publicly available parsers. 
        Unfortunately, organizations are required to monitor network traffic in order to identify the behavior of threat actors. 
        However, this dependency on proprietary and undocumented protocols, along with limited access to parsers, prevents the deployment of the necessary detection capabilities.

        \begin{table}[H]
            \begin{center}
            \caption{Available Network Parsers for ICS Protocols.\label{Table:11}}
                \begin{tabular}{|m{8em}|m{8em}|m{10em}|}
                    \hline
                    \textbf{Threat/vuln} & \textbf{Protocol} & \textbf{Parsers available} \\ \hline
                    \multirow{3}{*}{Industoryer} & IEC 61850 MMS & Yes \\
                     & IEC 60870-5-104 & Yes \\
                     & OPC DA & No \\ \hline
                    \multirow{3}{*}{Stuxnet} & S7Comm & Yes \\
                     & MS SQL(TDS) & Yes \\
                     & Profibus & No, but Profinet does \\ \hline
                    Triton & Tristation & No \\ \hline
                    Icsa-23-348-15 & PCOM & No \\ \hline
                    Icsa-23-193-01 & CIP & Yes \\ \hline
                \end{tabular}
            \end{center}
        \end{table}
        
        While proprietary and undocumented protocols were acceptable when vendors could assume only their own devices would operate within a closed ecosystem, the need to integrate network monitoring capabilities invalidates this assumption.
        Therefore, vendors must provide standard documentation and available parsers for network protocols to ensure asset owners can sufficiently detect threat actor behaviors in their environments.

        \item \textbf{Vulnerability advisories lack sufficient technical details to develop a detection of exploited vulnerability and lack rules/analytics to support that detection.}
        Current industry norms suggest that vendors are responsible for providing timely patches or updates to address product vulnerabilities, thereby mitigating the associated risks. However, it is well known that most asset owners struggle to perform patching in \ac{OT} environments due to shortages of \ac{OT} security personnel, limited maintenance periods, and high availability requirements. If a patch cannot be deployed, the organization may not have an alternative mechanism to mitigate that risk. It may only be able to monitor for any exploitation activity of this vulnerability. Unfortunately, vulnerability advisories constantly lacked an associated detection rule or analytic, and also lacked sufficient technical details necessary to devise an effective rule. Specifically, in the sampling of the nine vendor advisories reviewed for this study, five did not have mitigations other than patching, and seven lacked sufficient detail to support the detection, and only two included actual detection rules. Without sufficient technical details about vulnerabilities, asset owners cannot effectively develop their own detection rules or analytics, as this would likely require time- and resource-intensive reverse engineering of the device and associated protocols. Organizations are therefore primarily dependent on external security firms to perform these actions, but this ultimately presents additional costs to the organizations, both through the cost of the detection product and through the staff time dealing with false positives from an ineffective or imprecise rule. The two vulnerability advisories that contribute to associated detection rules demonstrate the feasibility of publishing these rules and should be acknowledged as an exemplar of leading industry practices.

        \item \textbf{Threat reports/advisories lack sufficient technical details or supporting technical artifacts across many procedure examples, constraining the development of detection rules or analytics.}
        Similar to the above challenge regarding the lack of technical details about vulnerabilities and associated detection rules or alerts, threat advisories do not consistently provide sufficient technical details or analytics/rules to enable the development of detections. 
        In the sampling of the 51 procedure examples identified for this study, at least 44 did not have sufficient technical detail to create a detection. 
        However, unlike vulnerability advisories, security firms or government agencies typically release threat advisories. 
        These organizations typically have limited access to information about threat activity due to constraints on the available data sources required for forensically analyzing the environments and devices. 
        Furthermore, as previously identified, detection vendors typically have limited information about vendor technologies, which may require additional reverse engineering to fully understand the threats’ technical actions and develop the required detections. 
        Therefore, this challenge should not be perceived as an inadequacy or deficiency of the organizations developing these reports, especially when the developing organization is producing this information to support their commercial product. 
        Instead, this is only intended to highlight the need for a more consistent approach/format for the structure and content of threat advisories. 
        While many reports document specific ATT\&CK \acp{TTP}, they lack specific technical artifacts or technical details necessary to detect that threat.
    \end{enumerate}
\section{Limitations and Implications}
    This study, while comprehensive in scope, has several limitations that should be considered when interpreting the findings.
    The first limitation is that our analysis primarily relies on MITRE ATT\&CK for ICS as the aggregated source of threat intelligence. 
    While ATT\&CK represents one of the most comprehensive publicly available knowledge bases of \ac{ICS} threats, this choice introduces potential coverage bias. 
    ATT\&CK may overrepresent well-documented, publicly disclosed incidents while underrepresenting threats that remain classified or proprietary. 
    If alternative sources such as sector-specific \acp{ISAC}, regional \acp{CERT}, or commercial threat intelligence were analyzed, we might observe different patterns in observable types and documentation quality.
    A second important limitation arises from the methodological constraints. While we validated the LLM-based extraction through manual review and correction of 103 observables, automated extraction may have missed implicit observables or misclassified complex ICS-specific artifacts. 
    The 87\% inter-rater reliability, while acceptable, indicates some subjective interpretation in observable classification.

\section{Conclusion}
    This study presents the first comprehensive, evidence-based analysis of the challenges associated with threat information sharing in industrial control systems. Through a deep analysis of threat information from three major attacks, a systematic examination of 196 procedure examples from ATT\&CK for ICS, and nine recent vulnerability advisories, we have empirically demonstrated four critical barriers that impede effective threat intelligence sharing in OT environments.

    Our findings reveal a fundamental misalignment between the information needed for \ac{ICS} threat detection and the capabilities of current sharing mechanisms. 
    The analysis reveals that 53\% of ICS-specific observables lack adequate \ac{STIX} representation, and 19\% have no \ac{STIX} representation, while 77\% of extracted artifacts lack sufficient technical detail for operational use. 
    These are not merely technical oversights but systemic issues stemming from the unique characteristics of \ac{OT} environment's proprietary protocols, specialized devices, and domain-specific operational constraints.

    The implications extend beyond technical standards. Our results suggest that effective \ac{ICS} threat intelligence sharing requires a paradigm shift in how the community approaches information collection, representation, and dissemination. 
    Vendors must recognize that security-by-obscurity, achieved through undocumented protocols, is no longer tenable when these same protocols become attack vectors. 
    Security researchers must balance the need for operational security with providing actionable technical details. 
    Standard bodies must evolve frameworks like \ac{STIX} to accommodate ICS-specific artifacts that do not fit traditional \ac{IT} paradigms.    

    Looking forward, this research establishes a baseline for measuring progress in \ac{ICS} threat intelligence sharing. 
    The methodology developed, combining manual analysis with automated extraction, provides a reproducible framework for ongoing assessment.
    As the \ac{ICS} threat landscape continues to evolve, regular application of this methodology can track improvements in information sharing quality and identify emerging gaps.

    The path forward requires coordinated action across multiple stakeholders. We recommend: (i) expansion of \ac{STIX} to include ICS-specific observable types identified in this study; (ii) industry adoption of standardized vulnerability disclosure templates that include detection rules; (iii) development of open-source protocol parsers for critical \ac{ICS} protocols; and (iv) establishment of sector-specific information sharing practices that balance operational security with actionable intelligence.

\section{Summary}
    This paper presents a data-driven analysis of threat information sharing challenges in industrial control systems, examining real-world threat intelligence and vulnerability advisories. 
    We analyzed 196 procedure examples across 79 techniques from the MITRE ATT\&CK for ICS framework and nine recent \ac{CISA} vulnerability advisories, identifying four key challenges:

    \begin{enumerate}
        \item \textbf{Insufficient \ac{STIX} Observable Support:} 72\% of ICS-specific threat observables cannot be adequately represented in current \ac{STIX} standards, with critical gaps in \ac{PLC} logic representation, ICS protocol specifics, and device-level artifacts.

        \item \textbf{Dependency on Proprietary Technologies:} Critical threat detection depends on understanding undocumented proprietary protocols, with 4 out of 9 identified \ac{ICS} protocols lacking publicly available parsers.

        \item \textbf{Inadequate Vulnerability Advisory Details:}Only 22\% of reviewed advisories provided detection rules, and 78\% lacked sufficient technical detail for developing custom detections.

        \item \textbf{Limited Actionable Threat Intelligence:} Only 23\% of threat procedure examples contained actionable technical details, constraining defenders' ability to develop effective detections.
    \end{enumerate}
    \begin{flushleft}
    Our systematic methodology, which combines manual analysis with automated \ac{LLM}-based extraction, provides a reproducible framework for assessing the quality of threat intelligence. The findings demonstrate that current information sharing practices, while improved from previous years, still fail to meet the operational needs of ICS defenders. Addressing these challenges requires coordinated efforts from vendors, researchers, and standards bodies to develop ICS-specific sharing mechanisms that balance security concerns with operational utility.
    \end{flushleft}
    
\bibliographystyle{unsrt}
\bibliography{references}
\section*{APPENDIX}
\subsection*{Observable Extraction Prompt}

The following prompt was used to systematically extract observables from ATT\&CK for \ac{ICS} procedure descriptions using GPT o3-mini:

\begin{mdframed}[style=promptbox]
\small
\textbf{System Prompt:} You are a helpful Cybersecurity assistant for identifying observables in Cyber Threat Intelligence text snippets.

\vspace{0.3cm}
\textbf{Task}
\begin{enumerate}
    \item You will receive a text snippet of a \ac{CTI} report from a user.
    \item Read the given snippet (plain text) carefully.
    \item Extract \textbf{every observable} (artifact) mentioned -- do \textbf{not} omit any.
    \item For code snippets, include the \textbf{full} code, including triple backticks.
    \item For each observable, output a JSON object with the exact fields listed in the \textbf{Response format} section.
\end{enumerate}

\vspace{0.3cm}
\textbf{Definitions}

\begin{enumerate}
    \item \textbf{Actionable Observable}
    \begin{itemize}
        \item \textit{Unique \& specific} $\rightarrow$ a deterministic IDS/YARA/SIEM rule could match it with low FP.
        \item \textit{Immediately operable} \textbf{as-is} (code snippet, exact URL, command, file name or path, API function) \textbf{or} after simple transform (e.g., Base64 decode, hash lookup, parameter substitution, memory dump).
        \item \textit{Searchable} and can drive automated response.
        \item If the observable meets these criteria $\rightarrow$ \texttt{artifact\_details} = \textbf{"Actionable"}.
    \end{itemize}

    \item \textbf{Described Observable}
    \begin{itemize}
        \item Has notable specifics, but still not unique enough for detection; non-searchable.
        \item If the observable meets this criteria and \textit{not} the Actionable Observable criteria $\rightarrow$ \texttt{artifact\_details} = \textbf{"Described"}.
    \end{itemize}

    \item \textbf{Mentioned Observable}
    \begin{itemize}
        \item A non-searchable observable, which doesn't stand in the Actionable Observable criteria, nor the Described Observable criteria.
        \item For such observables $\rightarrow$ \texttt{artifact\_details} = \textbf{"Mentioned"}.
    \end{itemize}

    \item \textbf{STIX Supported}
    This evaluates whether the observable is documented as \ac{STIX} 2.1 Cyber-observable Object
    \begin{itemize}
        \item \textbf{Full}: the observable's type exists in \ac{STIX} Cyber-Observable Objects.
        \item \textbf{Partial}: the observable does not map \textit{cleanly} to a first-class \ac{STIX} SCO, but can be approximated or expressed indirectly,
        \textbf{or} supported only via \texttt{x\_} custom properties or the generic \texttt{artifact} object.
        
        \item \textbf{No}: the observable isn't Fully \ac{STIX} supported nor Partially supported
    \end{itemize}

    \item \textbf{Proprietary Artifact}
    \begin{itemize}
        \item Open/Standard Technology
        \item Proprietary-Documented Technology
        \item Proprietary-Undocumented Technology
    \end{itemize}
\end{enumerate}

\vspace{0.3cm}
\textbf{Fields to produce for every observable}

\begin{tabular}{|p{3cm}|p{10cm}|}
\hline
\textbf{Field} & \textbf{Description} \\
\hline
\texttt{observable\_value} & Exact string (or faithful paraphrase). Escape any internal backticks. \\
\hline
\texttt{artifact\_details} & "Mentioned" | "Described" | "Actionable" based on the definitions above. \\
\hline
\texttt{data\_source} & Where it can be observed or collected (see cheat-sheet below). \\
\hline
\texttt{classification} & Short type label (e.g., "ICS Command", "URL", "Software/Tool"). \\
\hline
\texttt{STIX\_supported} & "Full: <STIX\_Object\_Name>" | "Partial: <STIX\_Object\_Name>" | "No". \\
\hline
\texttt{proprietary\_artifact} & "Open/Standard Technology" | "Proprietary-Documented Technology" | "Proprietary-Undocumented Technology". \\
\hline
\texttt{parser} & Known open-source/commercial parser name(s) for the data format, else \texttt{null} or "N/A" if not applicable. \\
\hline
\texttt{notes} & Any extra comments or context (Markdown allowed), or \texttt{null} if none. \\
\hline
\end{tabular}

\vspace{0.3cm}
\textbf{Common data\_source cheat-sheet}

Network traffic • Netflow • PCAP • DNS logs • Web proxy logs • Endpoint (EDR) logs • System logs (Windows Event, syslog) • ICS historian • PLC ladder logic • Firewall logs • Cloud API audit logs • Memory dump • None (if not observable via telemetry)

\vspace{0.3cm}
\textbf{Response format (return only this JSON)}

\begin{lstlisting}[language=json,basicstyle=\tiny\ttfamily,breaklines=true,frame=single]
{
  "observables": [
    {
      "observable_value": "<VAL>",
      "artifact_details": "Mentioned | Described | Actionable",
      "data_source": "<text>",
      "classification": "<one of allowed values>",
      "STIX_supported": "Full: <STIX_Object_Name> | Partial: <STIX_Object_Name> | No",
      "proprietary_artifact": "Open/Standard Technology | Proprietary-Documented Technology | Proprietary-Undocumented Technology",
      "parser": "<text>" | null | "N/A",
      "notes": "<text>" | null
    }
  ]
}
\end{lstlisting}
\end{mdframed}

\end{document}

%% file: Acronyms.tex
\begin{acronym}
\acro{ICS}{industrial control system}
\acro{STIX}{Structured Threat Information Expression}
\acro{AIS}{Automated Indicator Sharing}
\acro{DOE}{Department of Energy}
\acro{CRISP}{Cybersecurity Risk Information Sharing Program}
\acro{CyOTE}{Cybersecurity for the Operational Technology Environment}
\acro{KEV}{Known Exploitable Vulnerability}
\acro{IOC}{indicators of compromise}
\acro{LLM}{large language model}
\acro{SCO}{STIX Cyber-observable Object}
\acro{CTI}{Cyber Threat Intelligence}
\acro{CPS}{Cyber-Physical System}
\acro{OT}{Operational Technology}
\acro{IT}{ Information Technology}
\acro{NIS}{Network and Information Systems}
\acro{PLC}{Programmable Logic Controller}
\acro{CERT}{Cyber Emergency Response Teams}
\acro{TTP}{Tactics, Techniques and Procedures}
\acro{API}{Application Programming Interface}
\acro{OS}{Operating System}
\acro{C2}{Command and Control}
\acro{CRC}{Cyclic Redundancy Check}
\acro{DLL}{dynamic-link library}
\acro{IEC}{International Electrotechnical Commission}
\acro{SCADA}{Supervisory Control and Data Acquisition}
\acro{IOA}{Information Object Addresses}
\acro{PCAP}{Packet Capture}
\acro{CVE}{Common Vulnerabilities and Exposures}
\acro{CWE}{Common Weakness Enumeration}
\acro{ISAC}{Information Sharing and Analysis Center}
\acro{CISA}{Cybersecurity and Infrastructure Security Agency}

\end{acronym}